\documentstyle[a4,12pt,epsfig]{article}
\begin{document}

\begin{flushright}
\framebox{\bf hep-ph/0210276} \\
BIHEP-TH-2002-48
\end{flushright}

\vspace{0.4cm}

\begin{center}
{\large\bf Hierarchical Neutrino Masses and Large Mixing Angles \\
from the Fritzsch Texture of Lepton Mass Matrices} 
\end{center}

\vspace{0.5cm}

\begin{center}
{\bf Zhi-zhong Xing}
\footnote{Electronic address: xingzz@mail.ihep.ac.cn} \\
{\it CCAST (World Laboratory), P.O. Box 8730, Beijing 100080, China} \\
{\it and}
{\it Institute of High Energy Physics, Chinese Academy of Sciences, \\
P.O. Box 918 (4), Beijing 100039, China}
\footnote{Mailing address}
\end{center}

\vspace{2cm}

\begin{abstract}
We show that the Fritzsch texture of lepton mass matrices can naturally
lead to the bi-large flavor mixing pattern, if three neutrinos have a 
normal but weak mass hierarchy (typically, $m_1 : m_2 : m_3 \sim 1 : 3 :10$).
The effective mass of the tritium beta decay and that of the neutrinoless 
double beta decay are too small to be observable in this ansatz,
but CP violation at the percent level is allowed and 
could be measured in long-baseline neutrino oscillations.
\end{abstract}

\newpage

\framebox{\Large\bf 1} ~
The solar and atmospheric neutrino anomalies established recently in 
SNO \cite{SNO} and Super-Kamiokande \cite{SK} experiments are most
likely due to neutrino oscillations, a quantum phenomenon which can 
naturally happen if neutrinos are massive and lepton flavors are mixed. 
In the framework of three-neutrino oscillations, the mixing angles associated 
with both solar and atmospheric neutrino conversions are found to be 
surprisingly large ($\theta_{\rm sun} \sim 30^\circ$ \cite{Smirnov} and 
$\theta_{\rm atm} > 37^\circ$ \cite{Shiozawa}).
To understand the smallness of neutrino masses and the largeness of
flavor mixing angles, many phenomenological ans$\rm\ddot{a}$tze
of lepton mass matrices have been proposed \cite{Review}. An interesting
category of the ans$\rm\ddot{a}$tze take account of {\it texture zeros} 
of charged lepton and neutrino mass matrices in a given flavor basis, 
from which some nontrivial relations between flavor mixing angles and 
lepton mass ratios can be derived.

The present paper follows a similar idea to study the Fritzsch texture 
of lepton mass matrices and its consequences on neutrino masses, flavor mixing
and CP violation. In the quark sector, the Fritzsch ansatz \cite{F78}   
is partly successful to interpret the strong hierarchy of quark 
masses and that of flavor mixing angles. A number of authors have 
applied the same ansatz to the lepton sector \cite{Fritzsch}, 
in order to calculate the angles of lepton flavor mixing in terms of 
the masses of charged leptons and neutrinos. In most of those works, however,  
the unknown spectrum of neutrino masses was assumed to be 
strongly hierarchical, leading consequently to a small mixing angle 
for the solar neutrino oscillation. 

Can the Fritzsch texture of lepton mass matrices be incorporated with the 
bi-large flavor mixing pattern, which is remarkably favored by 
current data of solar and atmospheric neutrino oscillations? 
The answer may certainly be affirmative, if one gives
up the assumption that neutrino masses perform a strong hierarchy
as charged lepton masses. Nevertheless, this interesting 
possibility has not been carefully examined in the literature.

We carry out a careful analysis of the Fritzsch texture of lepton mass 
matrices, and find that it is able to predict the bi-large flavor mixing 
pattern if three neutrinos have a normal but weak mass hierarchy 
(typically, $m_1 : m_2 : m_3 \sim 1 : 3 : 10$).
The absolute values of three neutrino masses can then be determined
from the experimental data of atmospheric and solar neutrino oscillations. 
Both the effective mass of the tritium beta decay and that of the 
neutrinoless double beta decay are too small to be observable in this 
ansatz, but CP violation at the percent level is allowed and could be 
measured in the upcoming long-baseline neutrino oscillation experiments.
 
\framebox{\Large\bf 2} ~ 
Let us consider a simplest extension of the standard electroweak 
model \cite{BP}, in which the effective mass term of charged leptons 
($M_l$) and Majorana neutrinos ($M_\nu$) can be written as 
\begin{equation}
-{\cal L}_{\rm mass} \; =\; \overline{(e ~,~ \mu ~,~ \tau)^{~}_{\rm L}} ~
M_l \left (\matrix{e \cr \mu \cr \tau \cr} \right )_{\rm R} + ~
\frac{1}{2} ~ \overline{(\nu_e ~,~ \nu_\mu ~,~ \nu_\tau)^{~}_{\rm L}} ~
M_\nu \left (\matrix{\nu^{\rm c}_e \cr \nu^{\rm c}_\mu \cr 
\nu^{\rm c}_\tau \cr} \right )_{\rm R} + ~ {\rm h.c.} \; ,
\end{equation}
where $\nu^{\rm c}_\alpha \equiv C \overline{\nu}^{\rm T}_\alpha$ 
(for $\alpha = e, \mu, \tau$) with $C$ being the charge-conjugation
operator. While $M_\nu$ must be symmetric, $M_l$ is in general arbitrary.
Without loss of generality, one may arrange $M_l$ to be Hermitian 
through a suitable redefinition of the right-handed fields of charged
leptons \cite{FJ}
\footnote{It is also plausible to take $M_l$ to be symmetric 
in some extensions of the standard model. 
For instance, all fermion mass matrices are dictated to be symmetric 
in the SO(10) grand unified models \cite{F75,BW}.}.
The concrete structures of $M_l$ and $M_\nu$ are unfortunately unknown.
As pointed out in Ref. \cite{Branco}, an appropriate weak-basis
transformation allows us to put $M_l$ in the ``nearest-neighbor''
mixing form or the Fritzsch texture within the left-right symmetric 
models, but $M_\nu$ cannot simultaneously have the same texture.
It is therefore a phenomenological assumption that both $M_l$
and $M_\nu$ are of the Fritzsch texture in a specific flavor basis:
\begin{eqnarray}
M_l & = & \left ( \matrix{
{\bf 0}	& C_l	& {\bf 0} \cr
C^*_l	& {\bf 0} & B_l \cr
{\bf 0}	& B^*_l	& A_l \cr} \right ) \; ,
\nonumber \\
M_\nu & = & \left ( \matrix{
{\bf 0}	& C_\nu	& {\bf 0} \cr
C_\nu	& {\bf 0}	& B_\nu \cr
{\bf 0}	& B_\nu	& A_\nu \cr} \right ) \; .
\end{eqnarray}
We remark that the texture zeros of $M_l$ are just a special choice of the
weak basis, but those of $M_\nu$ are nontrivial and may stem from some 
underlying flavor symmetries (see, e.g., Refs. \cite{Review,F78,Fritzsch} for
detailed discussions). Taking account of the observed hierarchy
of charged lepton masses, $m_\tau \gg m_\mu \gg m_e$, one naturally
expects that $|A_l| \gg |B_l| \gg |C_l|$ holds in $M_l$. There is no 
unique limitation on the parameters of $M_\nu$, on the other hand,
because the mass spectrum of neutrinos has not been definitely determined
from the present experimental data
\footnote{The mass-squared differences extracted from 
solar and atmospheric neutrino oscillation data hint at
three possible patterns of the neutrino mass spectrum:
(a) normal hierarchy: $m_1, m_2 \ll m_3$; (b) inverted hierarchy:
$m_1 \approx m_2 \gg m_3$; (c) approximate degeneracy:
$m_1 \approx m_2 \approx m_3$.}.

Without loss of generality, one may take $A_l$ and $A_\nu$ to be
real and positive. Then only the off-diagonal elements of $M_l$ and
$M_\nu$ are complex. It is possible to express $M_l$ and $M_\nu$ as
\begin{equation}
M_l  = P^\dagger_l \overline{M}_l P_l \; , ~~
M_\nu = P^{\rm T}_\nu \overline{M}_\nu P_\nu \; ,
\end{equation}
where
\begin{equation}
\overline{M}_{l,\nu} \; =\; \left ( \matrix{
{\bf 0}  	& |C_{l,\nu}|   & {\bf 0} \cr
|C_{l,\nu}|  	& {\bf 0}   	& |B_{l,\nu}| \cr
{\bf 0}  	& |B_{l,\nu}|  	& A_{l,\nu} \cr} \right ) \; , 
\end{equation}
and 
\begin{eqnarray}
P_l & = & \left ( \matrix{
1 ~ & 0 & 0 \cr
0 ~ & e^{i\varphi^{~}_l} & 0 \cr
0 ~ & 0 & e^{i(\varphi^{~}_l + \phi^{~}_l)} \cr} \right ) \; ,
\nonumber \\
P_\nu & = & \left ( \matrix{
e^{i(\varphi^{~}_\nu - \phi^{~}_\nu)} & 0 & ~ 0 \cr
0 & e^{i\phi^{~}_\nu} & ~ 0 \cr
0 & 0 & ~ 1 \cr} \right ) \;
\end{eqnarray}
with $\phi^{~}_{l,\nu} \equiv \arg (B_{l,\nu})$ and 
$\varphi^{~}_{l,\nu} \equiv \arg (C_{l,\nu})$. The real symmetric matrices
$\overline{M}_l$ and $\overline{M}_\nu$ 
can be diagonalized by use of the following unitary transformations:
\begin{eqnarray} 
U^{\rm T}_l \overline{M}_l U_l & = & \left ( \matrix{
m_e & 0 & 0 \cr
0 & m_\mu & 0 \cr
0 & 0 & m_\tau \cr} \right ) \; , 
\nonumber \\
U^{\rm T}_\nu \overline{M}_\nu U_\nu & = & \left ( \matrix{
m_1 & 0 & 0 \cr
0 & m_2 & 0 \cr
0 & 0 & m_3 \cr} \right ) \; , 
\end{eqnarray}
where $(m_e, m_\mu, m_\tau)$ and $(m_1, m_2, m_3)$ are the physical masses
of charged leptons and neutrinos, respectively.   
The matrix elements of $U_l$ and $U_\nu$ depend on four mass ratios:
\begin{eqnarray}
x^{~}_l & \equiv & \frac{m_e}{m_\mu} \; , ~~~
x_\nu \; \equiv \; \frac{m_1}{m_2} \; ;
\nonumber \\
y^{~}_l & \equiv & \frac{m_\mu}{m_\tau} \; , ~~~
y_\nu \; \equiv \; \frac{m_2}{m_3} \; .
\end{eqnarray}
As the values of $m_e$, $m_\mu$ and $m_\tau$ have been determined
to a good degree of accuracy \cite{PDG}, we obtain
$x^{~}_l \approx 0.00484$ and $y^{~}_l \approx 0.0594$. The analytically 
exact results for nine elements of $U_l$ or $U_\nu$ are found to be
\begin{eqnarray}
U_{11} & = & + \left [ \frac{1-y}{(1+x)(1-xy)(1-y+xy)} \right ]^{1/2} \; , 
\nonumber \\
U_{12} & = & -i \left [ \frac{x(1+xy)}{(1+x)(1+y)(1-y+xy)} \right ]^{1/2} \; ,
\nonumber \\
U_{13} & = & + \left [ \frac{xy^3 (1-x)}{(1-xy)(1+y)(1-y+xy)} 
\right ]^{1/2} \; , 
\nonumber \\
U_{21} & = & + \left [ \frac{x(1-y)}{(1+x)(1-xy)} \right ]^{1/2} \; ,
\nonumber \\
U_{22} & = & +i \left [ \frac{1+xy}{(1+x)(1+y)} \right ]^{1/2} \; , 
\nonumber \\
U_{23} & = & + \left [ \frac{y(1-x)}{(1-xy)(1+y)} \right ]^{1/2} \; ,
\nonumber \\
U_{31} & = & - \left [ \frac{xy(1-x)(1+xy)}{(1+x)(1-xy)(1-y+xy)}
\right ]^{1/2} \; , 
\nonumber \\
U_{32} & = & -i \left [ \frac{y(1-x)(1-y)}{(1+x)(1+y)(1-y+xy)} 
\right ]^{1/2} \; , 
\nonumber \\
U_{33} & = & + \left [ \frac{(1-y)(1+xy)}{(1-xy)(1+y)(1-y+xy)} 
\right ]^{1/2} \; ,  
\end{eqnarray}
where we have omitted the index ``$l$'' or ``$\nu$'' for
simplicity. Note that $U_{i2}$ (for $i=1,2,3$) are imaginary,
and their nontrivial phases are due to the negative determinant of 
$\overline{M}_{l,\nu}$. Note also that two possibilities are allowed for 
the parameter space of $x_\nu$ and $y_\nu$ in Eq. (8)
\footnote{The extreme cases such as $x_\nu = 0$ or 1 and (or)
$y_\nu = 0$ or 1 are not taken into account, because they are
incompatible with current experimental data on
solar and atmospheric neutrino oscillations.}:
(a) $0 < x_\nu < 1$ and $0 < y_\nu < 1$; (b) $ x_\nu > 1$ 
and $y_\nu > 1$. Now that $x_\nu > 1$ is disfavored in respect 
of the large-mixing-angle MSW solution to the solar neutrino 
problem \cite{Smirnov}, we focus only on possibility (a) in
the following (namely, $m_1 < m_2 < m_3$).

The lepton flavor mixing matrix $V$ arises from 
the mismatch between the diagonalization of the charged lepton mass matrix 
$M_l$ and that of the neutrino mass matrix $M_\nu$.
In view of Eqs. (3) -- (6), we obtain 
$V = U^{\rm T}_l \left (P_l P_\nu \right ) U^*_\nu$, whose nine matrix
elements read explicitly as
\begin{equation}
V_{pq} \; = \; U^l_{1 p} U^{\nu *}_{1 q} e^{i\alpha} + 
U^l_{2 p} U^{\nu *}_{2 q} e^{i \beta} + U^l_{3 p} U^{\nu *}_{3 q} \; ,
\end{equation}
where the subscripts $p$ and $q$ run respectively over $(e, \mu, \tau)$
and $(1,2,3)$, and two phase parameters $\alpha$ and $\beta$ are defined
as $\alpha \equiv (\varphi^{~}_\nu - \varphi^{~}_l) - (\phi_\nu + \phi_l)$
and $\beta \equiv (\phi_\nu - \phi_l)$.
Note that an overall phase factor $e^{i(\varphi^{~}_l + \phi_l)}$ has been
omitted from the right-hand side of Eq. (9), since it has no contribution
to lepton flavor mixing and CP violation. 
It is obvious that $V$ consists of six parameters:
$x^{~}_l$, $y^{~}_l$, $x_\nu$, $y_\nu$, $\alpha$ and $\beta$. Among them, 
$x^{~}_l$ and $y^{~}_l$ are already known.
The other four free parameters can be constrained by current 
experimental data on neutrino oscillations.

\framebox{\Large\bf 3} ~
In the framework of three-neutrino oscillations, the solar and
atmospheric neutrino anomalies are approximately decoupled: they
are attributed, respectively, to $\nu_e \rightarrow \nu_\mu$ and 
$\nu_\mu \rightarrow \nu_\tau$ conversions. 
The corresponding neutrino mass-squared differences $\Delta m^2_{\rm sun}$ 
and $\Delta m^2_{\rm atm}$ are given by
\begin{eqnarray}
\Delta m^2_{\rm sun} & \equiv & m^2_2 - m^2_1 \; = \; m^2_2
\left (1 - x^2_\nu \right ) \; ,
\nonumber \\
\Delta m^2_{\rm atm} & \equiv & m^2_3 - m^2_2 \; = \; m^2_3
\left ( 1 - y^2_\nu \right ) \; .
\end{eqnarray}
The analyses of current SNO \cite{SNO} and Super-Kamiokande \cite{SK} data 
yield $\Delta m^2_{\rm sun} = (3.3 - 17) \times 10^{-5} ~ {\rm eV}^2$ 
\cite{Smirnov} and 
$\Delta m^2_{\rm atm} = (1.6 - 3.9) \times 10^{-3} ~ {\rm eV}^2$ 
\cite{Shiozawa} at the $90\%$ confidence level. Then the ratio
\begin{equation}
R_\nu \; \equiv \; \frac{\Delta m^2_{\rm sun}}{\Delta m^2_{\rm atm}}
\; = \; y^2_\nu ~ \frac{1 - x^2_\nu}{1 - y^2_\nu} \;
\end{equation}
lies in the range $(0.85 - 10) \times 10^{-2}$. On the other hand,
the mixing factors of solar and atmospheric neutrino oscillations are
related to the relevant matrix elements of $V$ in the following way:
\begin{eqnarray}
\sin^2 2 \theta_{\rm sun} & = & 4 |V_{e1}|^2 |V_{e2}|^2 \; ,
\nonumber \\
\sin^2 2 \theta_{\rm atm} & = & 4 |V_{\mu 3}|^2 
\left (1 - |V_{\mu 3}|^2 \right ) \; .
\end{eqnarray}
At the $90\%$ confidence level, $\tan^2 \theta_{\rm sun} = 0.30 - 0.58$
\cite{Smirnov} and $\sin^2 2\theta_{\rm atm} > 0.92$ \cite{Shiozawa}
have been extracted from current data. Note that the magnitude of $V_{e3}$ 
is well restricted by the CHOOZ reactor experiment on neutrino 
oscillations \cite{CHOOZ}, whose mixing factor is given by
\begin{equation}
\sin^2 2\theta_{\rm chz} \; \approx \; 4 |V_{e3}|^2
\left ( 1 - |V_{e3}|^2 \right ) \; .
\end{equation}
The present upper bound is $\sin^2 2\theta_{\rm chz} < 0.2$ or
$\theta_{\rm chz} < 13^\circ$ \cite{CHOOZ},
corresponding to the mass scale of atmospheric neutrino oscillations.

With the help of Eqs. (11) -- (13), the four unknown parameters 
$x_\nu$, $y_\nu$, $\alpha$ and $\beta$ in the Fritzsch texture of lepton
mass matrices can be determined from current experimental constraints on 
$R_\nu$, $\sin^2 2 \theta_{\rm sun}$, $\sin^2 2 \theta_{\rm atm}$ and
$\sin^2 2 \theta_{\rm chz}$. A careful numerical analysis shows that the phase
parameter $\beta$ must be around $180^\circ$ to assure 
$\sin^2 2\theta_{\rm atm} > 0.92$.
In comparison, the phase parameter $\alpha$ is not very restrictive.
Hence we simply fix $\beta = 180^\circ$ in our calculations, and typically take
$\alpha = (0^\circ, 90^\circ, 180^\circ)$ to illustrate the quantitative
dependence of four observables on $\alpha$. We present the allowed regions of 
$(x_\nu, y_\nu)$, $(\tan^2\theta_{\rm sun}, \sin^2 2\theta_{\rm atm})$ and
$(R_\nu, \sin^2 2\theta_{\rm chz})$ in Figs. 1, 2 and 3, respectively.
Some comments are in order.

(1) We demonstrate that the Fritzsch texture of lepton mass matrices 
is actually compatible with the present experimental data on solar and 
atmospheric neutrino oscillations. The allowed parameter space decreases 
with the change of $\alpha$ from $0^\circ$ to $180^\circ$. Typically, 
we obtain $x_\nu \sim 1/3$ and $y_\nu \sim (1/4 - 1/3)$ or 
\begin{equation}
m_1 : m_2 : m_3 \; \sim \; 1 : 3 : 10 \;\; .
\end{equation}
This result indicates that three neutrino masses are weakly hierarchical. In
contrast, three charged lepton masses 
($m_e : m_\mu : m_\tau \approx 1 : 207 : 3478$ \cite{PDG}) are strongly
hierarchical. 

(2) Given the numerical ranges of $x_\nu$ and $y_\nu$, the absolute values of
three neutrino masses can be calculated by use of Eq. (10) \cite{Xing02}:
\begin{eqnarray}
m_3 & = & \frac{1}{\sqrt{1 - y^2_\nu}} \sqrt{\Delta m^2_{\rm atm}} \;\; ,
\nonumber \\
m_2 & = & \frac{y_\nu}{\sqrt{1 - y^2_\nu}} \sqrt{\Delta m^2_{\rm atm}} 
\nonumber \\
& = & \frac{1}{\sqrt{1 - x^2_\nu}} \sqrt{\Delta m^2_{\rm sun}} \;\; ,
\nonumber \\
m_1 & = & \frac{x_\nu}{\sqrt{1 - x^2_\nu}} \sqrt{\Delta m^2_{\rm sun}} \;\; .
\end{eqnarray}
In view of the smallness of $x_\nu$ and $y_\nu$, we simplify Eq. (15)  
as $m_3 \approx \sqrt{\Delta m^2_{\rm atm}} ~$,
$m_2 \approx y_\nu \sqrt{\Delta m^2_{\rm atm}} 
\approx \sqrt{\Delta m^2_{\rm sun}} ~$, 
and $m_1 \approx x_\nu \sqrt{\Delta m^2_{\rm sun}} ~$. It is then 
straightfoward to arrive at $m_3 \approx (4.0 - 6.2) \times 10^{-2}$ eV,
$m_2 \approx (0.57 - 1.3) \times 10^{-2}$ eV, and
$m_1 \sim (1.9 - 4.3) \times 10^{-3}$ eV (for $x_\nu \sim 1/3$). 
More restrictive predictions for three neutrino masses will become available,
once the parameters of solar and atmospheric neutrino oscillations are
measured to a better degree of accuracy.

(3) Note that it is impossible to obtain $\sin^2 2\theta_{\rm atm} =1$ from
the Fritzsch texture of lepton mass matrices. The reason is simply that
$\theta_{\rm atm} > 40^\circ$ requires $y_\nu > 1/3$, which leads in turn
to $R_\nu > 0.1$ in this ansatz. In the leading-order approximation, we find 
\begin{equation}
|V_{\mu 3}| \; \approx \; \frac{\left |\sqrt{y_\nu (1 - x_\nu)} ~ e^{i\beta} 
- \sqrt{y^{~}_l} \right |}{\sqrt{1 + y_\nu}} \;\; 
\end{equation}
from Eqs. (8) -- (9). It becomes clear that $\beta = 180^\circ$ is most 
favored to make the magnitude of $V_{\mu 3}$ sufficiently large, such that 
the requirement $\sin^2 2\theta_{\rm atm} >0.92$ can be satisfied.

(4) As $x_\nu$ is close to $y_\nu$ in the allowed parameter space, we
have $R_\nu \approx y^2_\nu$ approximately. Hence 
$0.05 < R_\nu \leq 0.1$ is naturally anticipated. 
On the other hand, only small values of $\sin^2 2 \theta_{\rm chz}$ are
allowed in the Fritzsch texture of lepton mass matrices. More precise data
on $R_\nu$ and $\sin^2 2\theta_{\rm chz}$ may therefore provide a stringent 
test of this phenomenological ansatz.

Let us proceed to calculate the effective mass of the tritium beta decay
and that of the neutrinoless double beta decay:
\begin{eqnarray}
\langle m\rangle_e & \equiv & \sum^3_{i=1} \left ( m_i |V_{ei}|^2 \right )
\nonumber \\
& = & m_3 \left (x_\nu y_\nu |V_{e1}|^2 + y_\nu |V_{e2}|^2 + 
|V_{e3}|^2 \right ) \; ,
\nonumber \\
\langle m\rangle_{ee} & \equiv & 
\left | \sum^3_{i=1} \left ( m_i V^2_{ei} \right ) \right |
\nonumber \\
& = & m_3 \left | x_\nu y_\nu V^2_{e1} + y_\nu V^2_{e2} + 
V^2_{e3} \right | \; .
\end{eqnarray}
We present the allowed ranges of $\langle m\rangle_e$ and 
$\langle m\rangle_{ee}$, normalized by $m_3$, in Fig. 4. 
Given $m_3 \sim 5 \times 10^{-2}$ eV, 
$\langle m\rangle_e \sim 10^{-2}$ eV and 
$\langle m\rangle_{ee} \sim 10^{-3}$ eV are typically obtained.
Note that $\langle m\rangle_{ee} \ll \langle m\rangle_e$ holds in 
our ansatz, because the former involves significant cancellation
due to the existence of an extra phase (of $90^\circ$) 
associated with $V_{e2}$, as one can see from Eqs. (8) and (9).
It should be noted that our prediction for $\langle m\rangle_e$
is far below the proposed sensitivity of the KATRIN
experiment ($\sim 0.3$ eV \cite{K}), and that for $\langle m\rangle_{ee}$
is also below the expected sensitivity of next-generation experiments
for the neutrinoless double beta decay ($\sim 10$ meV to 50 meV \cite{B}).
Therefore it seems hopeless to detect both effects in practice.

How big can the strength of CP violation be in neutrino oscillations?
To answer this question, we calculate the Jarlskog invariant of CP violation 
$\cal J$ \cite{Jarlskog}, which is defined through the following equation:
\begin{equation}
{\rm Im} \left ( V_{a i} V_{b j} V^*_{a j} V^*_{b i} \right ) \; =\; 
{\cal J} \sum_{c, k} \left ( \epsilon_{a b c} \epsilon_{ijk} \right ) \; ,
\end{equation}
where the subscripts $(a, b, c)$ and $(i, j, k)$ run respectively over 
$(e, \mu, \tau)$ and $(1,2,3)$. We show the numerical changes of $\cal J$ 
with $|V_{e3}|$ in Fig. 5, where $\alpha = 90^\circ$ and $\beta = 180^\circ$ 
have been used. It is clear that the magnitude of $\cal J$ increases with that 
of $V_{e3}$. We obtain ${\cal J} \sim 1\%$ as a typical result in the allowed 
parameter space of $x_\nu$ and $y_\nu$. Thus leptonic CP-violating effects 
could be measured in a variety of long-baseline neutrino oscillation 
experiments \cite{LBL}.

\framebox{\Large\bf 4} ~
We have shown that the Fritzsch texture of lepton mass matrices is 
actually able to yield the bi-large flavor mixing pattern, if three 
neutrinos perform a normal but weak mass hierarchy. The absolute values of
three neutrino masses are approximately 
$m_1 \sim \sqrt{\Delta m^2_{\rm sun}}/3$,
$m_2 \sim \sqrt{\Delta m^2_{\rm sun}} ~$ and
$m_3 \sim \sqrt{\Delta m^2_{\rm atm}} ~$. The smallness of $m_i$ 
implies that there is little chance to observe the effective mass of the
tritium beta decay and that of the neutrinoless double beta decay. 
On the other hand, our ansatz predicts that the strength of leptonic CP
violation can be at the percent level. It is therefore possible to detect
CP-violating effects in the long-baseline experiments of
neutrino oscillations.

The weakly hierarchical spectrum of three neutrino masses must be a 
consequence of the weakly hierarchical texture of the neutrino mass
matrix $M_\nu$. To see this point more clearly, we calculate the
nonvanishing matrix elements of $\overline{M}_\nu$ in Eq. (4):
\begin{eqnarray}
A_\nu & = & m_1 - m_2 + m_3 \; ,
\nonumber \\
|B_\nu| & = & \left [ \frac{(m_1 - m_2) (m_2 - m_3) (m_1 + m_3)}
{m_1 - m_2 + m_3} \right ]^{1/2} \; ,
\nonumber \\
|C_\nu| & = & \left ( \frac{m_1 m_2 m_3}{m_1 - m_2 + m_3} \right )^{1/2} \; .
\end{eqnarray}
Taking the typical mass spectrum given in Eq. (14) as well as
$m_3 = 5 \times 10^{-2} ~ {\rm eV}$, we explicitly obtain
\begin{equation}
\overline{M}_\nu \; \sim \; 4 \times 10^{-2} ~ {\rm eV} \times 
\left ( \matrix{
{\bf 0}  & 0.24   & {\bf 0} \cr
0.24  	& {\bf 0}   & 0.55 \cr
{\bf 0}  & 0.55 & {\bf 1} \cr} \right ) \; .
\end{equation}
In comparison, the real charged lepton mass matrix $\overline{M}_l$ 
reads 
\begin{equation}
\overline{M}_l \; \approx \; 1.67 ~ {\rm GeV} \times 
\left ( \matrix{
{\bf 0}  & 0.0045   & {\bf 0} \cr
0.0045 	& {\bf 0}   & 0.26 \cr
{\bf 0}  & 0.26 & {\bf 1} \cr} \right ) \; .
\end{equation}
It becomes obvious that the mixing angle $\theta_{\rm sun}$ is
dominated by the $(1,2)$ subsector of $\overline{M}_\nu$, while
the mixing angle $\theta_{\rm atm}$ gets comparable contributions
from the $(2, 3)$ subsector of $\overline{M}_\nu$ and the
$(\mu, \tau)$ subsector of $\overline{M}_l$. This observation
would be useful for model building, from which some deeper understanding
of the Fritzsch texture or its variations could be gained. 

To conclude, the Fritzsch texture of lepton mass matrices is 
predictive and its predictions are compatible with current experimental
data on solar and atmospheric neutrino oscillations. A stringent test
of this ansatz will soon be available, in particular, in a variety of
long-baseline neutrino oscillation experiments.

\vspace{0.5cm}

The author would like to thank H. Fritzsch for very helpful discussions
and comments. This work was supported in part by the National Natural
Science Foundation of China.

\newpage

\begin{figure}[t]
\vspace{-2cm}
\epsfig{file=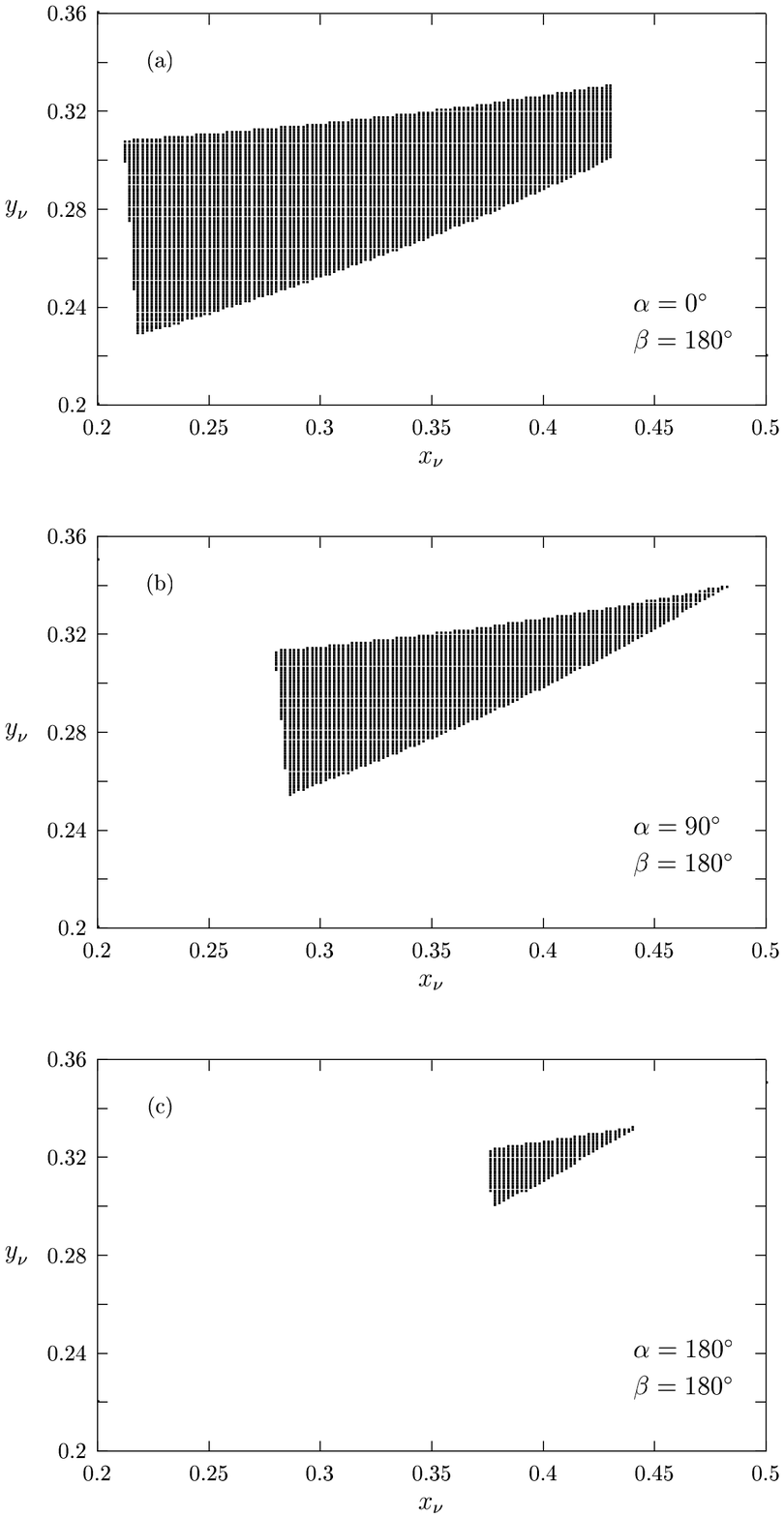,bbllx=-2.5cm,bblly=2cm,bburx=20cm,bbury=28cm,%
width=16cm,height=24cm,angle=0,clip=0}
\vspace{1cm}
\caption{Allowed regions of $x_\nu$ and $y^{~}_\nu$ for chosen values
of $\alpha$ and $\beta$.}
\end{figure}

\begin{figure}[t]
\vspace{-2cm}
\epsfig{file=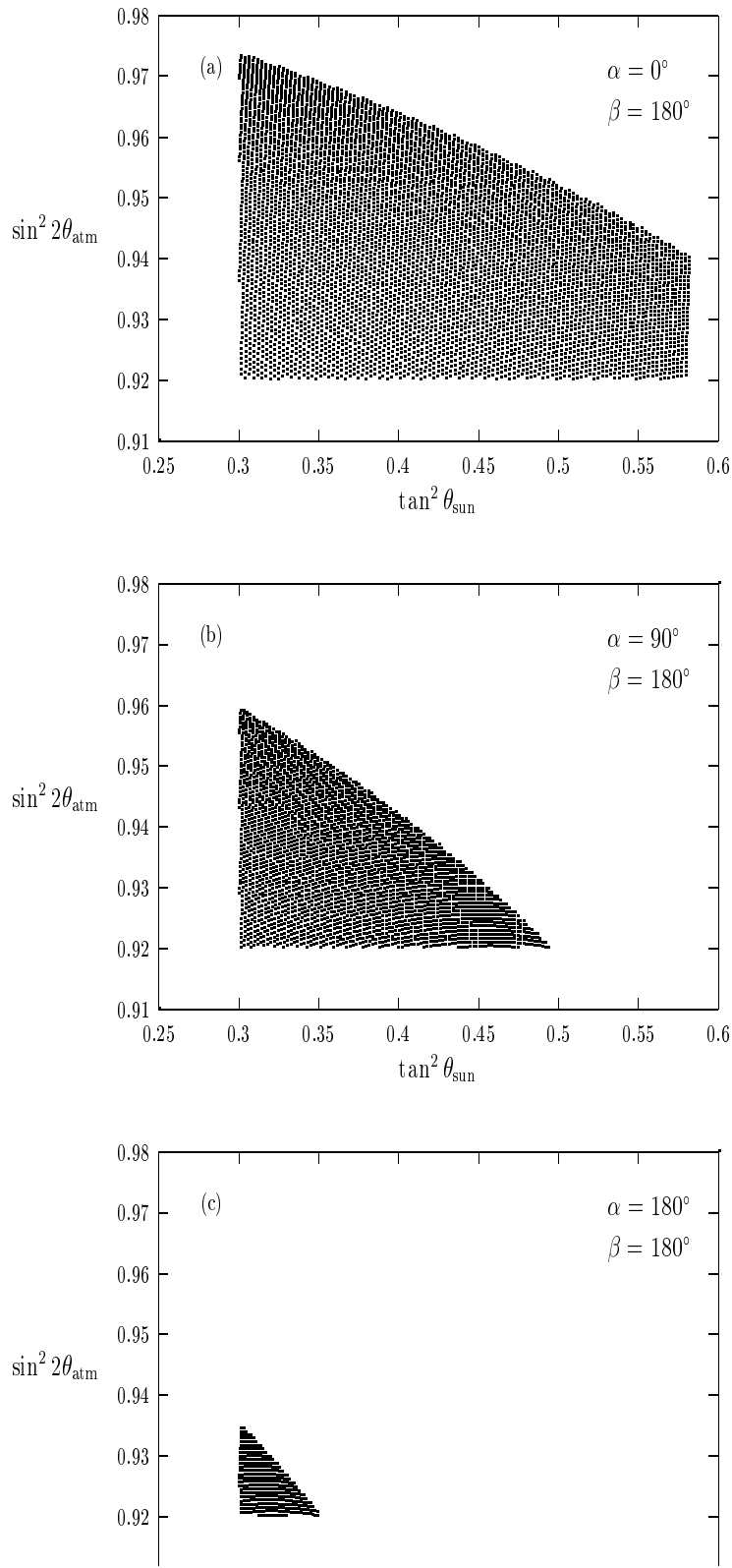,bbllx=-2.5cm,bblly=2cm,bburx=20cm,bbury=28cm,%
width=16cm,height=24cm,angle=0,clip=0}
\vspace{1cm}
\caption{Allowed regions of $\tan^2\theta_{\rm sun}$ and 
$\sin^2 2\theta_{\rm atm}$ 
for chosen values of $\alpha$ and $\beta$.}
\end{figure}

\begin{figure}[t]
\vspace{-2cm}
\epsfig{file=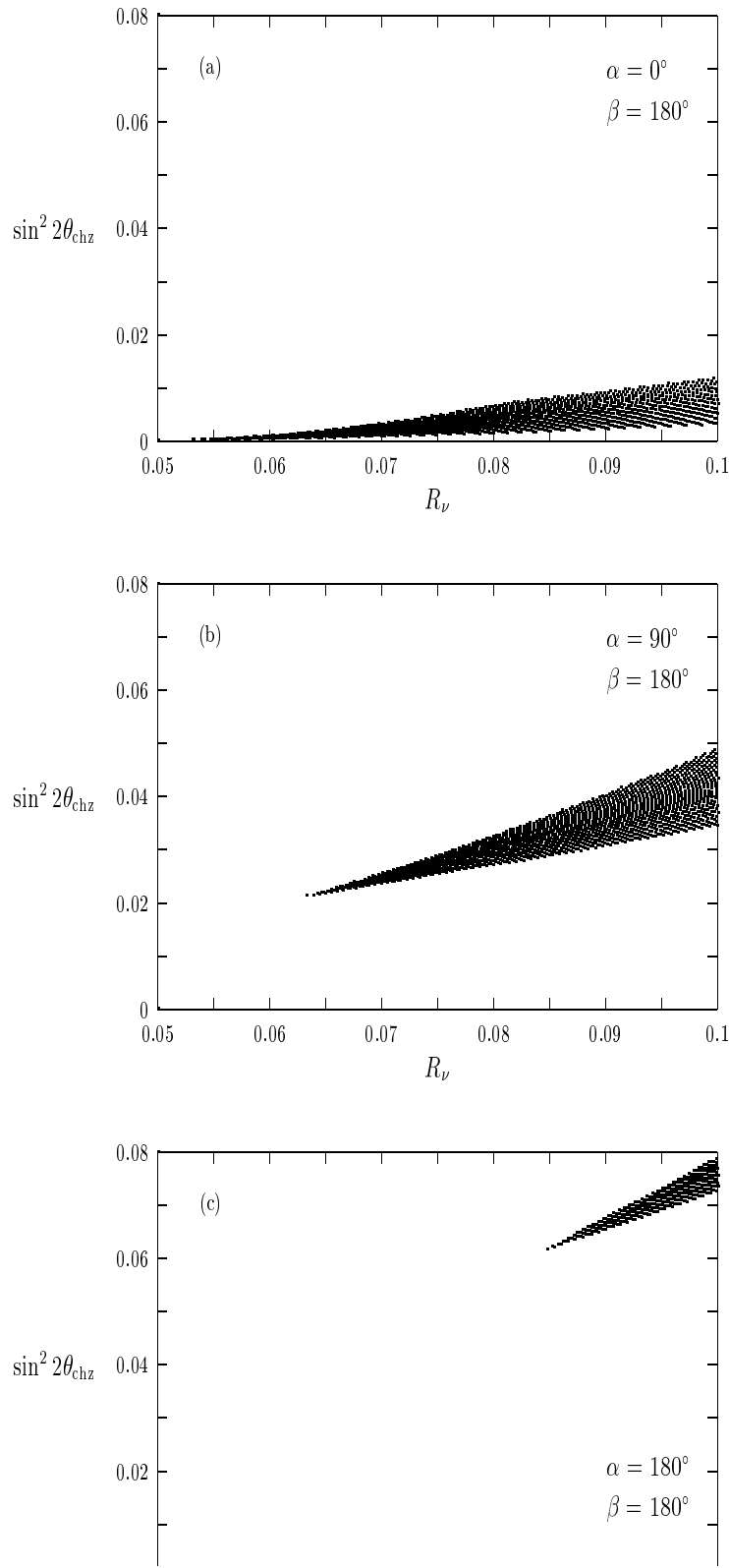,bbllx=-2.5cm,bblly=2cm,bburx=20cm,bbury=28cm,%
width=16cm,height=24cm,angle=0,clip=0}
\vspace{1cm}
\caption{Allowed regions of $R_\nu$ and $\sin^2 2\theta_{\rm chz}$ 
for chosen values of $\alpha$ and $\beta$.}
\end{figure}

\begin{figure}[t]
\vspace{-2cm}
\epsfig{file=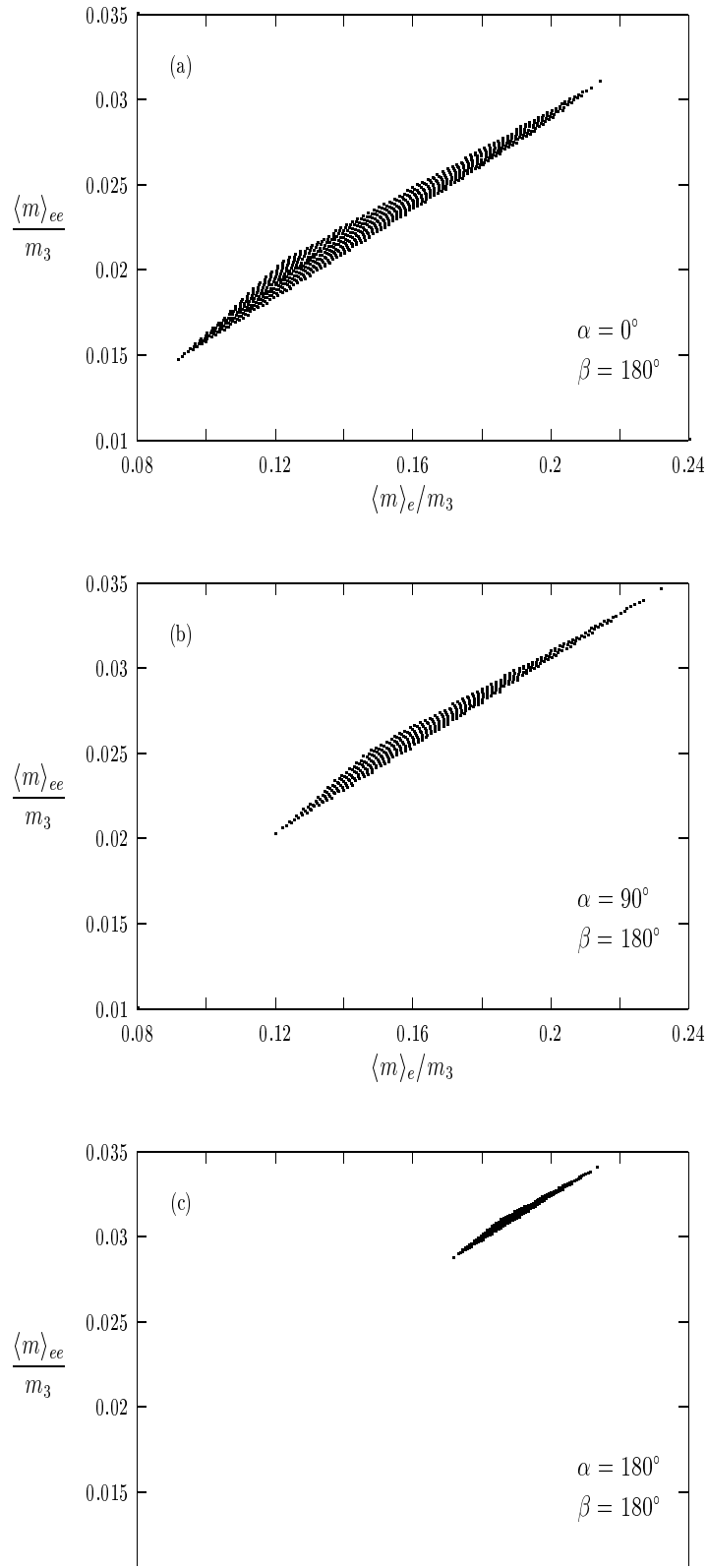,bbllx=-2.5cm,bblly=2cm,bburx=20cm,bbury=28cm,%
width=16cm,height=24cm,angle=0,clip=0}
\vspace{1cm}
\caption{Allowed regions of $\langle m\rangle_e/m_3$ and 
$\langle m\rangle_{ee}/m_3$ 
for chosen values of $\alpha$ and $\beta$.}
\end{figure}

\begin{figure}[t]
\epsfig{file=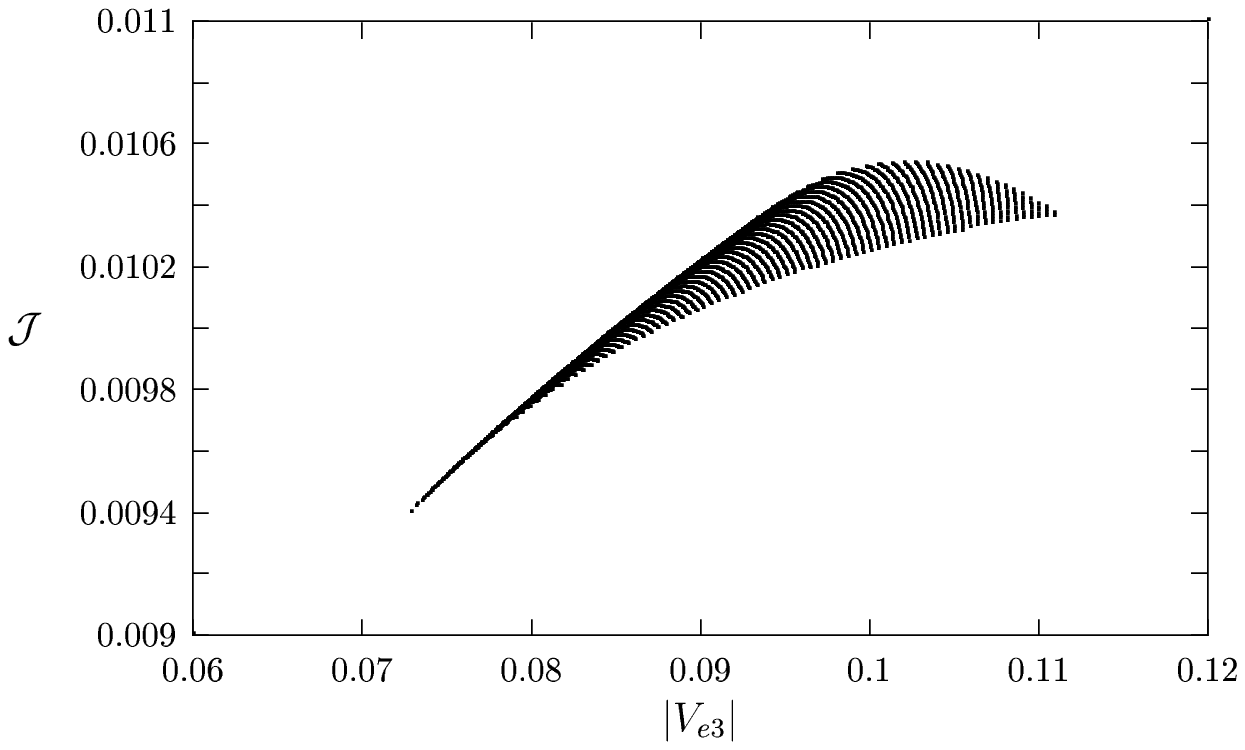,bbllx=-2.5cm,bblly=2cm,bburx=20cm,bbury=28cm,%
width=16cm,height=24cm,angle=0,clip=0}
\vspace{-14.3cm}
\caption{Allowed regions of $|V_{e3}|$ and $\cal J$ 
for the inputs $\alpha = 90^\circ$ and $\beta = 180^\circ$.}
\end{figure}


\newpage

\begin{thebibliography}{99}
\bibitem{SNO} SNO Collaboration, Q.R. Ahmad {\it et al.},
Phys. Rev. Lett. {\bf 87} (2001) 071301; {\bf 89} (2002) 011301; 
{\bf 89} (2002) 011302.

\bibitem{SK} Super-Kamiokande Collaboration,
Y. Fukuda {\it et al.}, Phys. Lett. B {\bf 467} (1999) 185;
S. Fukuda {\it et al.}, Phys. Rev. Lett. {\bf 85} (2000) 3999;
Phys. Rev. Lett. {\bf 86} (2001) 5651;
Phys. Rev. Lett. {\bf 86} (2001) 5656.

\bibitem{Smirnov} A.Yu. Smirnov, hep-ph/0209131; and references therein.

\bibitem{Shiozawa} M. Shiozawa (Super-Kamiokande Collaboration), talk given
at {\it Neutrino 2002}, Munich, May 2002.

\bibitem{Review} For recent reviews with extensive references, see:
H. Fritzsch and Z.Z. Xing, Prog. Part. Nucl. Phys. {\bf 45} (2000) 1;
G. Altarelli and F. Feruglio, hep-ph/0206077, to appear in
{\it Neutrino Mass} - Springer Tracts in Modern Physics, edited by
G. Altarelli and K. Winter (2002).

\bibitem{F78} H. Fritzsch, Phys. Lett. B {\bf 73} (1978) 317;
Nucl. Phys. B {\bf 155} (1979) 189.

\bibitem{Fritzsch} J.L. Hewett and T.G. Rizzo, 
Phys. Rev. D {\bf 33} (1986) 1519;
A.J. Davies and X.G. He, Phys. Rev. D {\bf 46} (1992) 3208;
M. Fukugita, M. Tanimoto, and T. Yanagida, 
Prog. Theor. Phys. {\bf 89} (1993) 263;
K.S. Babu and Q. Shafi, Phys. Lett. B {\bf 311} (1993) 172;
Y. Achiman and T. Greiner, Phys. Lett. B {\bf 329} (1994) 33.
For an application of the Fritzsch ansatz only to the charged lepton 
sector, see: C. Liu and J. Song, Phys. Rev. D {\bf 65} (2002) 057303.

\bibitem{BP} See, e.g., S.M. Bilenky and S.T. Petcov,
Rev. Mod. Phys. {\bf 59} (1987) 671; and references therein.

\bibitem{FJ} P.H. Frampton and C. Jarlskog, 
Phys. Lett. B {\bf 154} (1985) 421.

\bibitem{F75} H. Fritzsch and P. Minkowski, Annals Phys. {\bf 93} (1975) 193;
H. Georgi, in {\it Particles and Fields}, edited by C.E. Carlson
(AIP, NY, 1975), p. 575.

\bibitem{BW}  W. Buchm$\rm\ddot{u}$ller and D. Wyler,
Phys. Lett. B {\bf 521} (2001) 291;
Z.Z. Xing, Phys. Lett. B {\bf 545} (2002) 352.

\bibitem{Branco} G.C. Branco, L. Lavoura, and F. Mota,
Phys. Rev. D {\bf 39} (1989) 3443.

\bibitem{PDG} Particle Data Group, K. Hagiwara {\it et al.},
Phys. Rev. D {\bf 66} (2002) 010001.

\bibitem{CHOOZ} CHOOZ Collaboration, M. Apollonio {\it et al.},
Phys. Lett. B {\bf 420} (1998) 397;
Palo Verde Collaboration, F. Boehm {\it et al.},
Phys. Rev. Lett. {\bf 84} (2000) 3764.

\bibitem{Xing02} Z.Z. Xing, Phys. Rev. D {\bf 65} (2002) 077302;
Phys. Lett. B {\bf 530} (2002) 159;
Phys. Lett. B {\bf 539} (2002) 85.

\bibitem{K} V. Aseev {\it et al.}, talks given at the International
Workshop on Neutrino Masses in the Sub-eV Ranges, Bad Liebenzell, Germany,
January 2001; Homepage: http://www-ikl.fzk.de/tritium.

\bibitem{B} O. Cremonesi, hep-ph/0210007; and references therein.

\bibitem{Jarlskog} C. Jarlskog, Phys. Rev. Lett. {\bf 55} (1985) 1039;
H. Fritzsch and Z.Z. Xing, Nucl. Phys. B {\bf 556} (1999) 49; 
Phys. Rev. D {\bf 61} (2000) 073016.

\bibitem{LBL} See, e.g., A. Blondel {\it et al.},
Nucl. Instrum. Meth. A {\bf 451} (2000) 102;
C. Albright {\it et al.}, hep-ex/0008064;
D. Wark, plenary talk given at {\it ICHEP 2002}, Amsterdam, July 2002.
\end{thebibliography}
\end{document}